Specialized Field: Highway Engineering

Title: **Pavement Performance Evaluation Models for South Carolina**

Name: Rahman, Md Mostaqur (Corresponding Author)
Affiliation: Graduate Research Assistant
Department of Civil and Environmental Engineering
University of South Carolina
Address: 300 Main St
Columbia, SC 29208
USA
Telephone: (505) 730-1258
Fax: (803) 777-0670
E-mail: rahmanmm@cec.sc.edu

Name: Uddin, M. Majbah
Affiliation: Graduate Research Assistant
Department of Civil and Environmental Engineering
University of South Carolina
Address: 300 Main St
Columbia, SC 29208
USA
E-mail: muddin@cec.sc.edu

Name: Gassman, Sarah L.
Affiliation: Associate Professor
Department of Civil and Environmental Engineering
University of South Carolina
Address: 300 Main St
Columbia, SC 29208
USA
E-mail: gassman@cec.sc.edu

**Abstract**
This paper develops pavement performance evaluation models using data from primary and interstate highway systems in the state of South Carolina, USA. Twenty pavement sections are selected from across the state, and historical pavement performance data of those sections are collected. A total of 8 models were developed based on regression techniques, which include 4 for Asphalt Concrete (AC) pavements and 4 for Jointed Plain Concrete Pavements (JPCP). Four different performance indicators are considered as response variables in the statistical analysis: Present Serviceability Index (PSI), Pavement Distress Index (PDI), Pavement Quality Index (PQI), and International Roughness Index (IRI). Annual Average Daily Traffic (AADT), Free Flow Speed (FFS), precipitation, temperature, and soil type (soil Type A from Blue Ridge and Piedmont Region, and soil Type B from Coastal Plain and Sediment Region) are considered as predictor variables. Results showed that AADT, FFS, and precipitation have statistically significant effects on PSI and IRI for both JPCP and AC pavements. Temperature showed significant effect only on PDI and PQI ($p < 0.01$) for AC pavements. Considering soil type, Type B soil produced statistically higher PDI and PQI ($p < 0.01$) compared to Type A soil on AC pavements; whereas, Type B soil produced statistically higher IRI and PSI ($p < 0.001$) compared to Type A soil on . Using the developed models, local transportation agencies could estimate future corrective actions, such as maintenance and rehabilitation, as well as future pavement performances.

# 1. Introduction

The South Carolina Department of Transportation (SCDOT) is currently conducting the first phase of research on Mechanistic-Empirical Pavement Design Guide (MEPDG) local calibration for the state of South Carolina (SC), USA. The purpose of this research is to identify sources of data within SCDOT for calibration and to identify in-service pavement sections suitable for calibration studies. Selection of the hierarchical input level for each input parameter is necessary for pavement sections. The new MEPDG requires input data in four major categories: climate, traffic, materials, and pavement performance. The biggest challenge to use the performance data from SCDOT's specific pavement management system is the incompatibility of the SCDOT pavement performance data collection protocols with the new MEPDG distress identification protocol. Hence, there is a need for developing performance evaluation models for the SC pavements, by taking into account both local and MEPDG distress indices.

Pavement performance prediction is essential for rationally allocating resources at the network level (Meegoda and Gao, 2014), including resources for future maintenance and rehabilitation actions. Transportation agencies can save money by reducing the pavement deterioration prediction error (Madanat, 1993). To determine the future performance of pavements, the present condition of the pavement and the variables that control the pavement deterioration must be known. Pavement condition in SC is assessed by network level pavement roughness and surface distress data annually collected on the interstate and the primary highway systems by the SCDOT. Collected condition data includes roughness, rutting, fatigue cracking, transverse cracking, longitudinal cracking, raveling and patching. These condition data are used to determine different pavement performance indicators: Present Serviceability Index (PSI), Pavement Distress Index (PDI), Pavement Quality Index (PQI), and International Roughness Index (IRI). IRI and PSI are functions of roughness, PDI is a function of different distresses; and, PQI is a function of both pavement serviceability and distresses. Factors that affect pavement condition and performance indicators can be categorized into three groups: factors related to traffic, factors related to climate and factors related to material. The effect of these factors on pavement deterioration and performance indicators varies as described in a few previous literatures (Archilla and Madanat, 2001; Cooper et al., 2012; and Orobio and Zaniewski, 2011).

Truck traffic volume, climate (e.g., temperature, precipitation), and pavement structural condition have been shown to contribute most significantly to the deterioration of pavement (Meegoda and Gao, 2014). The magnitude and the number of wheel load passes is the main contributor to deteriorate the pavement surface (Isa et al., 2005). Usually medium truck loading is used to predict pavement deterioration in terms of annual average daily traffic (AADT). In South Carolina, about 10,000 large freight trucks (typically, weighing more than 10,000 pounds) traveled on the major interstates each day in 2007 (Uddin and Huynh, 2015). Moreover, SC pavements are exposed to extreme summer temperatures that average near 32 °C (90 °F) during the day and the precipitation is primarily in the form of rainfall that averages about 127 cm/year (50 inch/year) (NCEI, 2015). With the change of temperature and moisture content, pavement material characteristics for different pavement layers are prone to change (Nassiri and Bayat, 2013). If the deformations of one or more of the components in a typical pavement (e.g., surface layer, base layer and subgrade layer) are sufficiently large to cause cracking of the surfacing material, a pavement may be considered as failed (Seed et al., 1962). SC soils that serve as the subgrade layer can be divided into two regions separated by the geological fall line: the Blue Ridge/Piedmont Region and the Coastal Plain/Sediment Region. The soils in each region have different characteristics and thus are expected to have different impacts on pavement condition and performance indices.

Prior to MEPDG local calibration, a statistical study is required to assess the influence of various climatic, traffic and material inputs on pavement deterioration as related to pavement roughness and pavement distress indicators. Thus, the purpose of this study is to develop performance evaluation models using regression techniques for one of the MEPDG performance indicators: International Roughness Index (IRI), and three of the SCDOT pavement performance indices: PSI, PDI and PQI. To achieve this, five different design inputs are considered for the study: AADT, Free Flow Speed (FFS), precipitation, temperature, and soil type. These inputs are selected based on their importance on pavement performance and availability in the SCDOT database.

## 2. Literature Review

Pavement performance evaluation models have been developed for several states in the USA. These include the pavement performance models that were developed for the Delaware Department of Transportation (DelDOT) using the pavement inventory data of their pavement management system (Mills et al., 2012). The variables they considered were pavement age, geometry, functional class, type of overlay, pavement condition rating, and the annual average volume of traffic. Simple and multiple regression analysis were used to develop performance models for their pavements. Performance models for flexible pavement were developed for Georgia using regression technique (Kim and Kim, 2006). The researchers found that linear regression models are effective to forecast pavement performance if Average Annual Daily Traffic (AADT) is considered as a predictor. Gulen et al. (2001) developed regression models to predict the performance of pavements in Indiana, where they considered pavement roughness as the response variable, and pavement age and AADT as predictor variables. Performance models were also developed using regression techniques for Minnesota pavements (Prozzi and Madanat, 2004). A network level pavement performance model was developed using 20 years of historical pavement condition data for approximately 19,000 highway sections maintained by the New York State Department of Transportation (NYSDOT) (DeLisle et al., 2003). In a study by the Michigan Department of Transportation (MDOT), pavement distress data was used to assess the impact of construction (smoothness, early completion of construction, and nuclear density) on pavement performance (Chang et al., 2001). Pavement performance model was also developed using Pavement Condition Rating for North Carolina Department of Transportation (NCDOT) (Chan et al., 1997); and an overall distress index, a structural index with roughness index for North Dakota Department of Transportation (NDDOT) (Johnson and Cation, 1992).

In addition to aforementioned studies in the USA, some other countries have developed performance models for their respective pavement systems: Malaysia (Isa et al., 2005), Portugal (Ferreira et al., 2010), Canada (Hong and Wang, 2003), and New Zealand (Henning et al., 2004). Isa et al. (2005) used regression techniques to develop pavement performance models for federal roads of Malaysia. Ferreira et al. (2010) tested two pavement performance models, the AASHTO model and the Nevada model for Portugal through the use of the strategic evaluation tool (SET) based on deterministic segment-linked optimization model and solved by a method developed using generic algorithm method. A simple probabilistic approach was developed for Ontario pavement based on nonhomogeneous continuous Markov chain by Hong and Wang (2003). In New Zealand, data from 63 Long Term Pavement Performance (LTPP) sites was used to calibrate the pavement deterioration models currently used on the state highway network (Henning et al., 2004).

Different statistical and regression techniques have been used to develop pavement evaluation models and to study the effects of different factors on pavement performance. Thyagarajan et al. (2010) studied the critical input parameters of MEPDG to investigate the effect of variability in key input parameters. The influence of project specific input uncertainties were evaluated on predicted pavement performance and distresses. They found Tornado plots and extreme tail analysis are useful statistical tools that can assist design engineers to identify the relative importance of input parameters and the effect of their variability on design reliability. Salama et al. (2006) investigated the effect of different axle and truck types on flexible pavement damage. Condition evaluation models were developed for Distress Index (DI), Ride Quality Index, and rutting. A relative comparison of different variables was carried out using simple, multiple and stepwise regression technique. An auto regression approach for predicting pavement DI was developed, in other study, with limited data (Ahmed et al., 2010). Gulen et al. (2001) also used regression techniques to develop improved performance prediction models. IRI was used as a response variable, while the age of the pavement and the current AADT were predictor variables. The data from their randomly selected road test sections did not yield statistically strong models. Xu et al. (2014) used linear regression and artificial neural networks to predict the deterioration of Wheel Path Cracking (WPC) over a one year period. The extent and severity of WPC along with age and AADT were used as input variables in the study. An empirical comparison of nine representative statistical pavement performance models was conducted by Chu and Durango-Cohen (2008) using serviceability data from the AASHTO road test. The purpose of the study was to understand the effect of different statistical assumptions and estimation techniques on the models predictive capabilities. Recently, Gupta et al. (2012) performed a critical review of the literature related to flexible pavement performance models. The paper presented a detailed review of various pavement performance models to examine the roles of factors related to pavement materials, environmental conditions, and type of traffic and volume of traffic. They concluded from other literatures that age and traffic are the most important variables to predicting pavement distress. Moreover, climate factors affect the structural properties of the pavements which are responsible for the deterioration of the pavements. Since these factors are uncertain in nature and vary from place to place, they considered them as important in analyzing the performance of pavement.

A minimum of 20 pavement test sections was recommended by Baus and Stires (2010) for calibrating and validating distress predictions. Hence, 20 pavement sections were selected from 15 counties in SC to serve as a representative sample—14 Asphalt Concrete (AC) sections of average length 5.3 miles and 6 Portland Cement Concrete (PCC) pavement sections of average length 5.8 miles. In the state of South Carolina, the mostly used PCC pavement type is Jointed Plain Concrete Pavement (JPCP) and all the 6 selected PCC sections are JPCP. None of the selected pavement sections are Continuously Reinforced Concrete Pavement or Jointed Reinforced Concrete Pavement.

**3. Objectives of Study**
The objectives of this study are to:
1. Develop performance evaluation models for AC pavements and JPCP using multiple regression techniques for different distress indicators: PSI, PDI, PQI, and IRI.

2. Investigate the effect of different variables (AADT, FFS, precipitation, temperature, and soil type) on AC and JPCP pavement performances.
   3. Compare AC and JPCP pavement performance for two unbound materials: soil type A from the Blue Ridge and Piedmont Region, and soil type B from Coastal Plain and Sediment Region

**4. Pavement Sections, Data, and Variables**
4.1 Pavement Sections Selection
Table 1 lists the selected pavement sections with their location, pavement type, surface course type and thickness, base course type and thickness, pavement length, and date of construction. Fig. 1 shows the location of the selected pavement sections. To select the in-service sections, the following guidelines were in consideration.
   1. The pavement sections are primary or interstate routes located in Coastal Plain and Piedmont Regions in SC.
   2. Both flexible and rigid pavements with typical layer configuration and material selection, including traditional and new materials, are included.
   3. Different service times for different types of pavements are included.
   4. Priority is given to the initially selected sections with historical data, including climate, materials, traffic, and performance data.
   5. Selected sections are not overlaid or rehabilitated, and are suitable for MEPDG local calibration.

Table 1. Selected Pavement Sections

| County | Location | Type | Surface Course Thickness (in.) | Base Course Type | Base Course Thickness (in.) | Length (miles) | Date of Construction |
|---|---|---|---|---|---|---|---|
| Aiken | I 520 | JPCP | 11 | AA + GAB | 1.5 + 8 | 5.35 | 7/25/2008 |
| Beaufort | US 278 | AC | 3.6 | AA + GAB | 3.2 + 6 | 1.56 | 3/13/1998 |
| Charleston | SC 461 | AC | 5.7 | AA + SAB | 2.7 + 8 | 2.48 | 5/21/1996 |
| Charleston | I 526 | JPCP | 11 | CSM + CMS | 6 + 6 | 2.39 | 6/25/1991 |
| Chester | SC 9 | AC | 6.1 | GAB | 8 | 7.12 | 10/1/1999 |
| Chesterfield | SC 151 | AC | 3.9 | AA + Sand Clay | 2.7 + 8 | 5.36 | 12/15/1999 |
| Fairfield | I 77 | JPCP | 10 | LC + CMS | 6 + 6 | 14.17 | 10/21/1980 |
| Florence | SC 327 | AC | 6.9 | Macadam | 8 | 5.09 | 2/25/1992 |
| Florence | US 301 | AC | 3.8 | GAB + CMS | 8 + 6 | 2.38 | 9/30/2003 |
| Georgetown | US 521 | AC | 3.8 | GAB + CMS | 8 + 6 | 4.07 | 6/1/2003 |
| Greenville | I 385 | AC | 16.6 | CSM | 6 | 7.65 | 8/28/2000 |
| Greenville | I 85 | AC | 3.9 | AA | 7.7 | 1.00 | 8/31/2005 |
| Horry | SC 22 | AC | 3.8 | AA + GAB | 5.5 + 8 | 24.35 | 10/12/2001 |
| Horry | SC 31 | AC | 3.8 | AA + GAB | 2.7 + 8 | 3.98 | 1/31/2005 |
| Laurens | SC 72 | AC | 3.6 | AA | 6.8 | 5.99 | 3/1/2002 |
| Lexington | S 378 | JPCP | 9 | GAB | 6 | 1.47 | 11/1/2001 |
| Orangeburg | US 321 | AC | 5.6 | GAB | 6 | 6.17 | 7/1/2004 |
| Pickens | SC 93 | AC | 3.4 | AA | 5.8 | 1.34 | 4/10/2001 |

| Spartanburg | SC 80 | JPCP | 10 | GAB | 5 | 3.30 | 6/1/2000 |
| Spartanburg | I 85 | JPCP | 12 | AA + CMS | 4 + 6 | 6.29 | 6/11/1997 |

Note: I, US, and SC represent Interstate highways, United States routes, and South Carolina routes, respectively. AC = Asphalt Concrete, JPCP = Jointed Plain Concrete Pavement, AA = Asphalt Aggregate Base, GAB = Graded Aggregate Base, SAB = Stabilized Aggregate Base, CSM = Cement Stabilized Macadam, CMS = Cement Modified Subbase, LC = Lean Concrete.

### 4.2 Performance Data

Historic performance data for the selected pavement sections were collected using the pavement viewer of SCDOT's Integrated Transportation Management System (ITMS). Available performance data for the past 10 years were collected and summarized. The data included four main performance measures: (1) PSI, (2) PDI, (3) PQI, and (4) IRI. A total of 160 data points, representing each performance indicator, were collected for the 20 pavement sections. The number of samples collected for the AC pavements was 103, and that of the JPCP pavements was 57. Descriptive statistics of the numerical variables are presented in Table 2. The performance indicators are described next, along with their value ranges.

Table 2. Descriptive Statistics of the Numerical Variables in the Evaluation Models

| Variable | Unit | AC pavements | | | JPCP pavements | | |
|---|---|---|---|---|---|---|---|
| | | Min | Max | Std. dev. | Min | Max | Std. dev. |
| PSI | – | 2.6 | 4.7 | 0.3 | 2.9 | 4.1 | 0.3 |
| PDI | – | 1.9 | 4.7 | 0.6 | 4.1 | 5.0 | 0.2 |
| PQI | – | 2.0 | 4.3 | 0.5 | 3.7 | 4.5 | 0.2 |
| IRI | inch/mile | 43.9 | 112.9 | 15.2 | 50.5 | 137.4 | 20.2 |
| $log_{10}$AADT | veh/d | 3.4 | 5.0 | 0.5 | 3.8 | 4.9 | 0.4 |
| FFS | mph | 39.7 | 71.2 | 7.9 | 47.6 | 75.6 | 7.7 |
| Precipitation | mm | 31.8 | 68.5 | 8.5 | 31.7 | 54.7 | 6.2 |
| Temperature | °F | 54.1 | 66.7 | 3.4 | 59.2 | 68.3 | 2.7 |

#### 4.2.1 Present Serviceability Index (PSI)

PSI represents the riding quality of the pavement and is calculated from the mean IRI. SCDOT uses the following equation developed by Paterson (1986) for estimating PSI.

$$PSI = 5.0e^{(-0.002841 \times IRI)} \quad (1)$$

where IRI is in inch/mile and PSI is a dimensionless index ranging from 0 to 5; 5 represents perfect condition and 0 represents failed condition. In the SCDOT pavement management system, a newly constructed pavement is assigned a PSI value of 4.5.

#### 4.2.2 Pavement Distress Index (PDI)

PDI describes the observed surface distresses for PCC pavements and observed surface distress with mean rut depth for AC pavements. For PCC pavements observed surface distress includes: punchouts, spalling, pumping, patching, transverse cracking, longitudinal cracking, faulting, and surface deterioration. For AC pavements observed surface distress includes: raveling, fatigue cracking, patching, transverse cracking, and longitudinal cracking. To determine PDI, SCDOT uses the following equation (PMS, 1990).

$$PDI = 5.0 - ADV \quad (2)$$

where ADV is the adjusted distress value. Description of the ADV can be found in different literatures (e.g., Wang, 2002). Newly constructed pavements are assumed to be distress free, meaning that the ADV is initially zero. Therefore, those pavements are assigned a PDI of 5.0. PDI ranges from 0 to 5, where 5 represents perfect condition and 0 represents failed condition. PDI is a dimensionless index similar to PSI.

4.2.3 Pavement Quality Index (PQI)
PQI is the combination of PSI and PDI, which represents the overall condition of the pavement. SCDOT determines PQI using following equation (PMS, 1990).

$$PQI = 1.158 + 0.138 \times PSI \times PDI \qquad (3)$$

where PQI is a dimensionless index and ranges from 0 to 5. In the SCDOT pavement management system, newly constructed pavement is assigned a PQI value of 4.3.

4.2.4 International Roughness Index (IRI)
IRI is an index for roughness measurement obtained by road meters installed on vehicles or trailers. In SC, IRI values are derived from wheel path profiles obtained using non-contacting inertial profilers. Typically, data readings are taken at 0.10 mile intervals and then averaged (Baus and Hong, 2004). IRI values less than 170 inch/mile are acceptable and any IRI value less than 95 inch/mile indicates good roughness condition of the pavement (FHWA, 2004; Shahin, 2005).

4.3 Predictor Variables
In this study, effects of the following five predictor variables are investigated: (1) Annual Average Daily Traffic (AADT), (2) Free flow speed (FFS), (3) Precipitation, (4) Temperature, and (5) Soil type. Variables (1) to (4) are numerical, and variable (5) is categorical.

4.3.1 Annual Average Daily Traffic (AADT)
AADT is the average daily traffic on a roadway section for all days of the week during a period of one year, expressed in vehicle per day (veh/d). The number of repeated traffic is solely responsible for the load related pavement distresses. AADT data were collected for the pavement sections for the period from 2005 to 2014.

4.3.2 Free Flow Speed (FFS)
FSS affects the pavement roughness and surface friction. The following equations (Dowling, 1997) were used to calculate the FFS of the pavement sections.

$$FFS = (0.88 * Link\ Speed\ Limit + 14); for\ speed\ limit > 50\ mph \qquad (4)$$

$$FFS = (0.79 * Link\ Speed\ Limit + 12); for\ speed\ limit \leq 50\ mph \qquad (5)$$

4.3.3 Precipitation

Moisture content is an environmentally driven variable that can affect the pavement layer properties, such as degradation of material quality, loss of bond between layers and softening of the subgrade layer (ARA, 2004). Hence, precipitation could affect the pavement performances. The mean annual precipitations of the pavement sections were taken from their corresponding counties, found from the National Climate Data Center database (NCEI, 2015), for the years 2005 to 2014.

4.3.4 Temperature

Temperature is another environmental factor, which affect pavement performance. For the selected pavements sections, temperature information was collected for their respective counties from the National Climate Data Center database (NCEI, 2015). Specifically, yearly mean temperature data were collected for the years 2005 to 2014.

4.3.5 Soil Type

Subgrade soil strength for the selected pavement sections is currently not available; therefore, different soil types have been chosen as an alternative. Two types of soils are selected: Type A and Type B. SC soils can be divided into two regions separated by the geological fall line as shown in Fig. 1: (i) Upstate Area or Blue Ridge and Piedmont Region (Type A), and (ii) Coastal Plain and Sediment Region (Type B) (SCDOT, 2010).

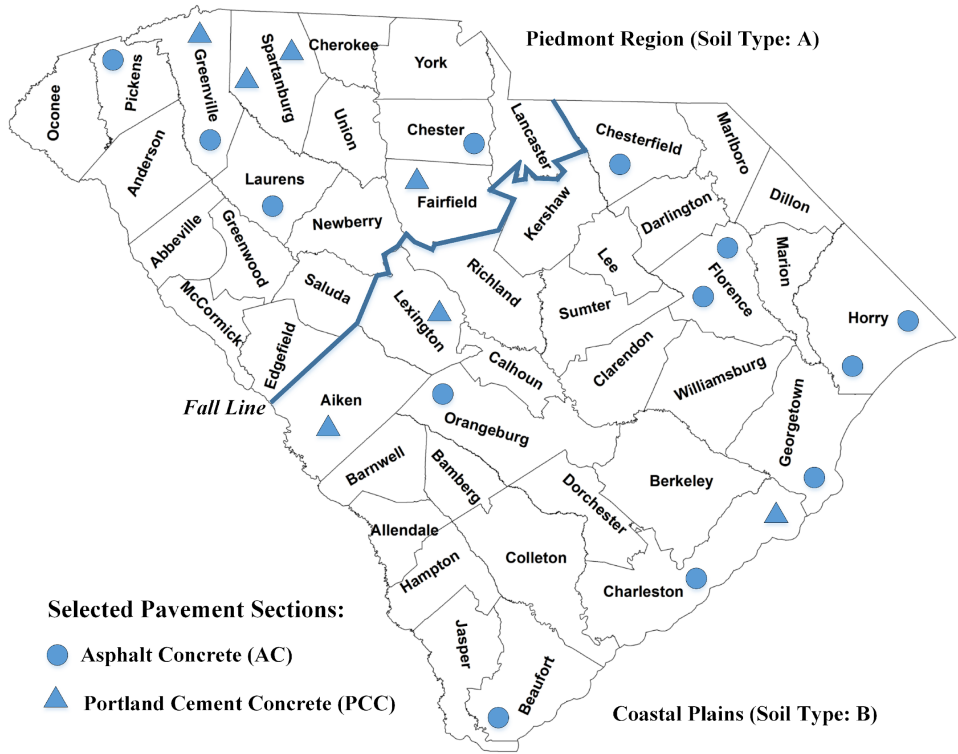

Fig. 1. Selected Pavement Sections with Counties

Type A soils are described as micaceous clayey silts and micaceous sandy silts, clays, and silty soils in partially drained condition; Type B soils include fine sand that is difficult to compact. In terms of AASHTO classifications, Type B soils are primarily A-1 to A-4 and Type A soils are predominately A-5 or higher (Pierce et al., 2011). The AASHTO system classifies soils into eight groups: A-1 through A-8 where A-1 to A-3 are granular soils, A-4 to A-7 are fine grained soils, and A-8 represents organic soils (AASHTO, 2008).

## 5. Methodology

To formulate pavement performance evaluation models, multiple linear regression analysis was conducted. Specifically, using multiple linear regression analysis, for performance measures/indicators, two separate models were formed for each performance indicator. The first model describes the effects of predictor variables on the indicators if the pavement is AC, and the second model describes the effects if the pavement is JPCP. The results of the models will be compared with each other and the best model will be suggested for the performance evaluation. A brief description on multiple linear regression and its assumptions are provided next.

### 5.1 Multiple Linear Regression

A linear regression model that contains more than one predictor/independent variable is called a multiple linear regression model. It takes into account the effect of all specified predictor/independent variables at the same time. Suppose the response variable $Y$ is quantitative and at least one predictor variable $X_i$ is quantitative, then the multiple linear regression models have the following form.

$$Y = \beta_0 + \beta_1 X_1 + \beta_2 X_2 + \cdots + \beta_i X_i$$

(6)

where $Y$ = response variable (e.g., IRI, PDI, PQI, and PSI); $\beta_0$ = intercept; $\beta_i$ = coefficients; and $X_i$ = predictor variables (e.g., $log_{10}AADT$, Temperature, Precipitation, FFS, and Soil Type). The intercept $\beta_0$ defines the value of $Y$ when all $X_i$'s are 0. The regression coefficient $\beta_k$ represents the change in the mean response corresponding to a unit change in $X_k$ when all other $X_i$'s are held constant, $k \in i$.

### 5.1.1 Assumptions

Multiple linear regression analysis makes several key assumptions. The principal assumptions (Keith, 2015) are described here.

1. Dependent and independent variables are linearly related through regression coefficients. An appropriate transformation of the variable must be incorporated in the model if there is non-linearity.
2. Each observation should be drawn independently from the population. This means that the errors for each observation are independent from those of others.
3. There must be equal variance of errors across all levels of the independent variables, which refer to as homoscedasticity.

4. The errors are normally distributed. This assumption is only vital in case of small samples.
5. There is little or no multicollinearity in the data. Multicollinearity occurs when several independent variables correlate at high levels with one another.

5.1.2 Standardized Regression Coefficients

To compare the relative importance of different predictor variables, standardized coefficients values are often utilized. Standardized coefficients are determined by converting all variables into Z scores, which in turn, convert the distribution mean to zero and standard deviation to one. The standardized multiple linear regression is specified as the following.

$$Y' = \beta'_1 X'_1 + \beta'_2 X'_2 + \cdots + \beta'_i X'_i \tag{7}$$

$$Y' = Z_y = \frac{\bar{Y} - Y}{\sigma_y} \tag{8}$$

$$X'_i = Z_{x_i} = \frac{\bar{X}_i - X_i}{\sigma_{x_i}} \tag{9}$$

where $Y'$ = standardized response variable; $\beta'_i$ = standardized coefficients; $X'_i$ = standardized predictor variables; $\bar{Y}$ = average value of response variable; $\sigma_y$ = standard deviation of response variable; $\bar{X}_i$ = average value of predictor variables; and $\sigma_{x_i}$ = standard deviation of predictor variables.

5.2 Analysis

The statistical analysis was started with a bivariate analysis to examine the Pearson intercorrelation among a distress indicator (e.g., PSI, IRI, and PDI) and the predictor variables. Bivariate analysis also involved the assessment of multicollinearity of the predictor variables. To do this, a variance inflation factor (VIF) was introduced. VIF measures how much the variance of a coefficient is increased due to multicollinearity, and a VIF ≥ 10 indicates a serious multicollinearity problem (Neter and Wasserman, 1996). Next, a multiple regression analysis was conducted to predict the values of response variable based on the value of predictor variables. Then, unstandardized and standardized coefficients for each predictor variables were analyzed to determine precisely the level of change in the response variable accounted for by a change in the predictor variable. The overall $R^2$ and adjusted $R^2$ of the regression model were calculated to assess the percentage of the variance in the distress indicator that was explained by the predictor variables. The aforementioned procedure was followed for each distress indicator and both pavement types. The procedure was repeated 8 times. All analyses were performed in IBM SPSS Statistics software (v 12).

## 6. Results

Results of the Pearson intercorrelation analysis of the PSI with the predictor variables, in case of AC pavements, are presented in Table 3. The correlation between PSI and predictor variables are found as low to large, with the Pearson correlation values ($r$) ranging from 0.05 to 0.58. FFS is the strongest related predictor of PSI ($r = 0.58$, $p < 0.01$). The table also shows that some of the predictor variables have strong correlations with each other. For instance, AADT is strongly correlated with Precipitation ($r = 0.39$, $p < 0.01$) and with Temperature ($r = -0.62$, $p < 0.01$). In contrast, the correlation between FFS and Temperature, and Soil type and Precipitation are found as low. Lastly, VIF values of predictors suggest that there is no serious multicollinearity problem in the data. Similar outputs are found from the Pearson intercorrelation analysis of the other distress indicators with the predictor variables. For the sake of brevity, those outputs are not presented here. However, all of the correlation and VIF values are within acceptable range (VIF < 10).

Performance evaluation models for PSI, PDI, PQI, and IRI are reported in Table 4 through Table 7, respectively. Each performance indicator has two different models: one for AC pavements and another one for JPCP. Each model shows different statistical results from the analyses; which include unstandardized regression coefficients ($\beta$), standardized regression coefficients ($\beta'$), coefficient of determination ($R^2$), and overall model significance (F-test). In the analysis, soil type B is considered as the reference soil type. Eight evaluation models were fitted; and each model was found overall statistically highly significant after the F-test, except for the PDI model for AC pavements ($p < 0.01$).

Table 3. Correlations for PSI and Predictor Variables, and VIF (AC Pavements)

|  | Pearson correlation ($r$) | | | | | | VIF |
|---|---|---|---|---|---|---|---|
|  | PSI | Precipitation | Temperature | $log_{10}$AADT | FFS | Soil type |  |
| PSI | 1.00 |  |  |  |  |  |  |
| Precipitation | -0.13 | 1.00 |  |  |  |  | 1.40 |
| Temperature | 0.05 | -0.27** | 1.00 |  |  |  | 4.58 |
| $log_{10}$AADT | -0.04 | 0.39** | -0.62** | 1.00 |  |  | 2.09 |
| FFS | 0.58** | 0.18 | -0.04 | 0.34** | 1.00 |  | 1.55 |
| Soil type | -0.20* | -0.01 | -0.81** | 0.43** | -0.26** | 1.00 | 4.41 |

*$p < 0.05$; **$p < 0.01$

Table 4. Effects of Predictor Variables on PSI

|  | AC pavements | | JPCP pavements | |
|---|---|---|---|---|
| Predictors | $\beta$ | $\beta'$ | $\beta$ | $\beta'$ |
| Intercept | 4.397*** |  | 6.229*** |  |
| Precipitation | -0.006* | -0.211 | -0.005* | -0.115 |
| Temperature | -0.019 | -0.249 | -0.021 | -0.202 |
| $log_{10}$AADT | -0.151** | -0.301 | -0.809*** | -1.057 |
| FFS | 0.021*** | 0.684 | 0.038*** | 1.044 |
| Soil type (A vs B) | -0.047 | -0.094 | -0.560*** | -1.018 |
| $R^2$ | 0.442 | | 0.868 | |
| Adjusted $R^2$ | 0.413 | | 0.855 | |
| Overall model significance | $F(5, 94) = 14.913$*** | | $F(5, 50) = 65.865$*** | |

*$p < 0.05$; **$p < 0.01$; ***$p < 0.001$

The effects of different independent variables on PSI for AC pavements and JPCP are shown in Table 4. In the PSI model for AC pavements, FFS, AADT and precipitation were found to have statistically significant effects on PSI. FFS showed positive effects on PSI, β = 0.021; $p < 0.001$. In contrast, AADT poses negative effects on PSI, β = -0.151; $p < 0.01$. Precipitation also showed negative effects on PSI, β = 0.006; $p < 0.05$. FFS showed higher absolute standardized regression coefficient (β' = 0.684) than AADT (β' = -0.301) or precipitation (β' = -0.211). This indicates that FFS has more importance to explain PSI of AC pavements than AADT or precipitation. The model was overall statistically significant, $F_{(5, 94)} = 14.913$; $p < 0.001$, and the model explained 44.2% of total variation in PSI ($R^2 = 0.442$).

For JPCP, four independent variables showed statistically significant effects on PSI. Among those variables, FFS showed positive effects on PSI (β = 0.038), whereas AADT (β = -0.809), precipitation (β = -0.005) and soil types (β= -0.560) showed negative effects on PSI. In JPCP, AADT, FFS and soil types showed the highest absolute standardized regression coefficient (β' = -1.057, β' = 1.044 and β' = -1.018 respectively). This means, among all the variables considered, these three have more importance to explain PSI of JPCP. Statistically significant unstandardized regression coefficient of soil type (β = -0.560) indicates that soil type A acts significantly to attain lower PSI measure than soil type B for JPCP. Like AC pavements, JPCP also showed overall statistically significant model, $F_{(5, 50)} = 65.865$; $p < 0.001$. Moreover, the model fit is very satisfactory with explaining 86.8% of total variation in PSI ($R^2 = 0.868$). AADT, FFS and precipitation showed statistically significant effects on PSI for both AC and JPCP models. Several literatures also found reduction in PSI due to traffic (e.g., Lu et al., 1974; Mikhail and Mamlouk, 1999).

Table 5 shows the effects of different predictor variables on PDI for AC pavements and JPCP. In the PDI model for AC pavements, FFS, temperature and soil types were found to have statistically significant effects on PDI. FFS showed positive effects on PDI, β = -0.033; $p < 0.001$. Temperature and soil type (A vs B) also showed positive effects on PDI (β= 0.104, β = 0.787, respectively). The model was overall moderately statistically significant, $F_{(5, 95)} = 3.938$; $p < 0.001$, and the model explained 17.2% of total variation in PDI, ($R^2 = 0.172$). Hence, the selected variables of this study are less capable of explaining PDI than PSI ($R^2 = 0.442$) for AC pavements. Results of PDI model agree with Hasan et al. (2015). They found that the mean annual temperature has a great influence on pavement distresses.

Table 5. Effects of Predictor Variables on PDI

| Predictors | AC pavements | | JPCP pavements | |
| --- | --- | --- | --- | --- |
| | β | β' | β | β' |
| Intercept | -4.565 | | 7.576*** | |
| Precipitation | 0.000 | -0.006 | -0.007* | -0.191 |
| Temperature | 0.104** | 0.577 | -0.012 | -0.155 |
| $log_{10}$AADT | -0.160 | 0.131 | -0.443*** | -0.751 |
| FFS | 0.033*** | 0.428 | 0.003 | 0.107 |
| Soil type (A vs B) | 0.787** | 0.641 | -0.114 | -0.271 |
| $R^2$ | | 0.172 | | 0.639 |
| Adjusted $R^2$ | | 0.128 | | 0.604 |

| Overall model significance | F(5, 95) = 3.938** | F(5, 51) = 18.071*** |

*$p < 0.05$; **$p < 0.01$; ***$p < 0.001$

For JPCP, precipitation and AADT showed statistically significant effects on PDI. Both variables showed negative impact on PDI ($\beta$ = -0.007 and $\beta$= -0.443 respectively). Unlike AC pavements, JPCP did not show any significant effect of temperature on PDI, $\beta$ = -0.012; $p > 0.05$. JPCP showed an overall statistically significant model, F (5, 51) = 18.071; $p < 0.001$. Moreover, the model satisfactorily explained 63.9% of the total variation in PDI ($R^2$ = 0.639).

Table 6 shows the effects of different predictor variables on PQI, which is a function of PSI and PDI (Eq. (2)). For AC pavements, FFS, temperature and soil type showed statistically significant effects on PQI. The model was overall statistically significant, F (5, 95) = 4.591; $p < 0.001$, and the model explained 19.5% of the total variation in PDI, ($R^2$ = 0.195). For JPCP, precipitation, AADT, FFS and soil type showed statistically significant effects on PQI. Unlike AC pavements, JPCP did not show any significant effect of temperature. JPCP showed overall statistically significant model, F (5, 51) = 27.470; $p < 0.001$. Moreover, the model explained 72.9% of total variation in PQI ($R^2$ = 0.729).

The regression results for IRI models are found in Table 7. For both AC and JPCP, precipitation, AADT and FFS showed statistically significant influence on IRI. Precipitation and AADT showed positive impact on IRI while FFS showed negative impact on IRI. In addition, soil type showed statistically significant influence on IRI for JPCP. This is similar to the findings of Al-Mansour et al. (1994). Like AADT, they found traffic count had a significant influence on pavement roughness. Moreover, another study also reported a negative effect of vehicle operational speed on IRI (Li et al., 2014), and Wang et al. (2014) found that pavement roughness has a very small effect on FFS. Temperature did not have a significant influence on IRI for either type of pavement, which agrees with the results from Hasan et al. (2015). Results also indicate that unlike AC pavements, soil type has a significant effect on IRI for JPCP.

Table 6. Effects of Predictor Variables on PQI

| Predictors | AC pavements | | JPCP pavements | |
|---|---|---|---|---|
| | $\beta$ | $\beta'$ | $\beta$ | $\beta'$ |
| Intercept | -2.683 | | 6.882*** | |
| Precipitation | -0.002 | -0.037 | -0.006* | -0.199 |
| Temperature | 0.076** | 0.530 | -0.015 | -0.200 |
| $log_{10}$AADT | 0.145 | -0.150 | -0.491*** | -0.889 |
| FFS | 0.029*** | 0.479 | 0.011** | 0.423 |
| Soil type (A vs B) | 0.604** | 0.620 | -0.220* | 0.557 |
| $R^2$ | 0.195 | | 0.729 | |
| Adjusted $R^2$ | 0.152 | | 0.703 | |
| Overall model significance | F(5, 95) = 4.591*** | | F(5, 51) = 27.470*** | |

*$p < 0.05$; **$p < 0.01$; ***$p < 0.001$

Table 7. Effects of Predictor Variables on IRI

| Predictors | AC pavements | | JPCP pavements | |
|---|---|---|---|---|
| | $\beta$ | $\beta'$ | $\beta$ | $\beta'$ |
| Intercept | -4.876 | | -116.797* | |
| Precipitation | 0.474* | 0.190 | 0.379* | 0.118 |
| Temperature | 1.852 | 0.296 | 1.648 | 0.225 |
| $log_{10}$AADT | 14.686** | 0.349 | 57.820*** | 1.053 |
| FFS | -1.789*** | -0.669 | -2.666*** | -1.032 |
| Soil type (A vs B) | 0.800 | 0.019 | 39.592*** | 1.003 |
| $R^2$ | | 0.407 | | 0.859 |
| Adjusted $R^2$ | | 0.376 | | 0.845 |
| Overall model significance | F(5, 97) = 13.315*** | | F(5, 50) = 61.109*** | |

*$p < 0.05$; **$p < 0.01$; ***$p < 0.001$

6.1 Model Comparison

Comparisons of measured and estimated distress indicator values for the AC pavements are presented in Fig. 2 and for the JPCP in Fig. 3. The figures illustrate how the measured distress indicators are consistent with the model estimates for each data record. For AC pavements, the figures show that the estimates from the models are generally well matched with the measured distress values, although those for PDI and PQI show more outliers than the other two distress indicators. For JPCP, the figures indicate a good match between model estimates and measured distress values.

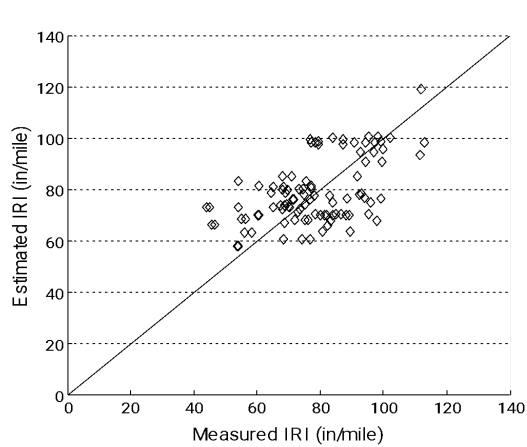
(a) IRI

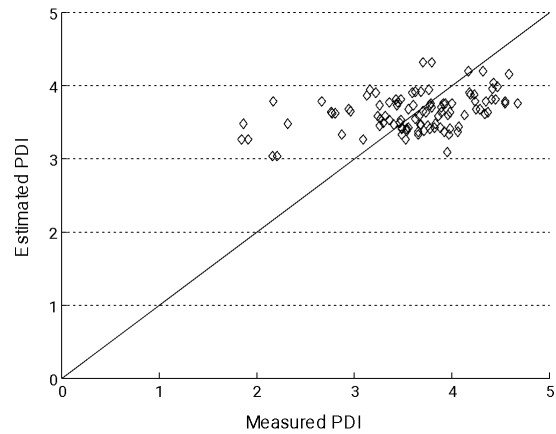
(b) PDI

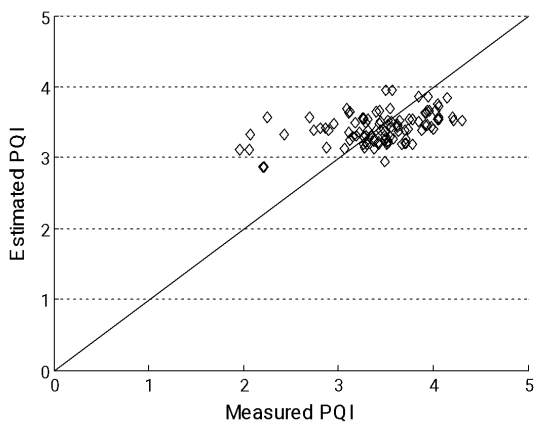
(c) PQI

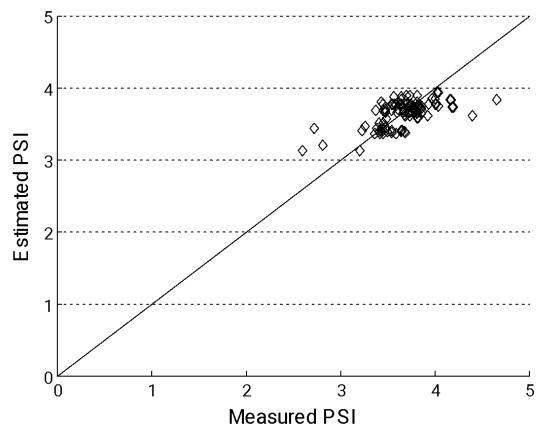
(d) PSI

(c) PQI                                                    (d) PSI

Fig. 2. Comparison of Estimated and Measured Performance Indicators for AC Pavements

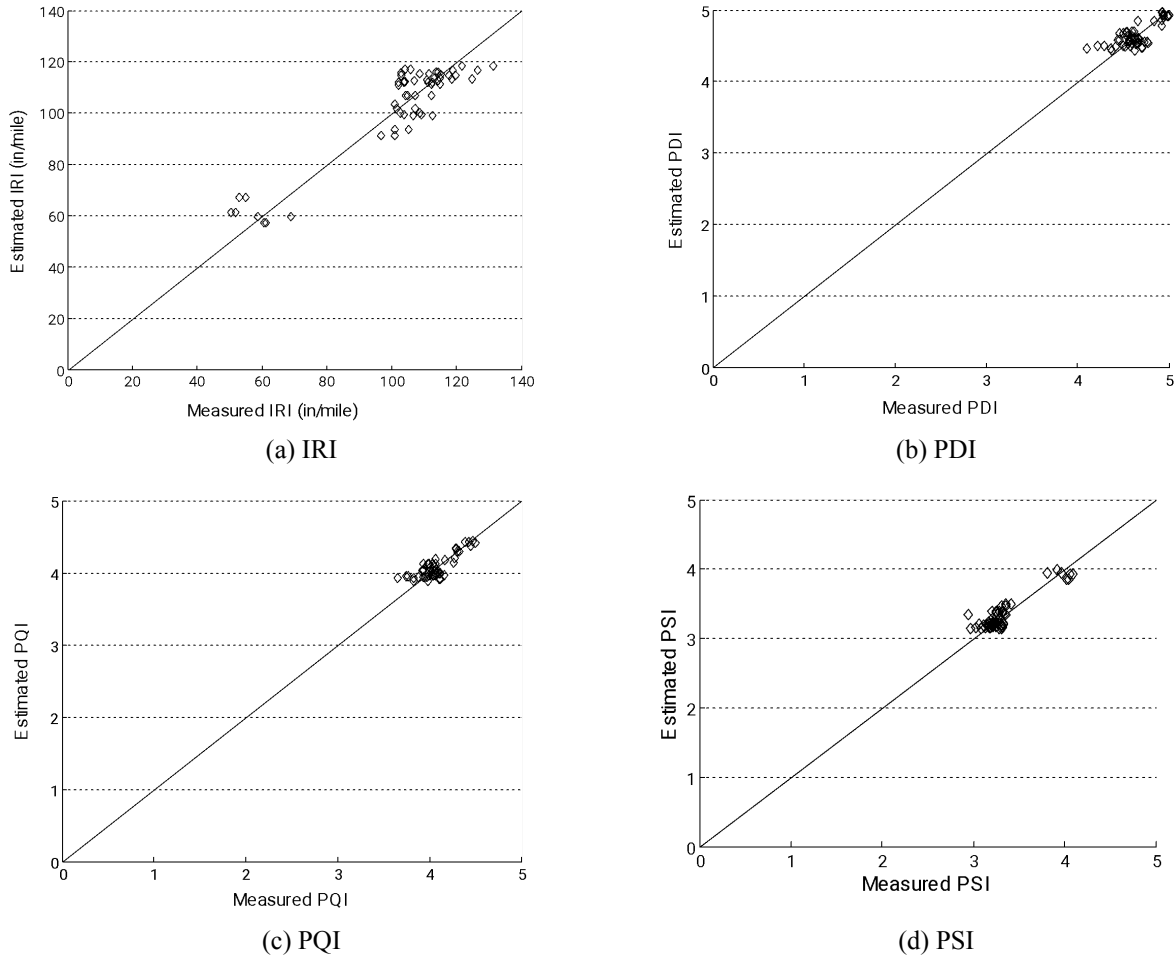

(a) IRI                                                    (b) PDI

(c) PQI                                                    (d) PSI

Fig. 3. Comparison of Estimated and Measured Performance Indicators for JPCP

Table 8 compares the mean values of measured and estimated distress indicators for both AC and JPCP. It can be seen that the estimated mean values are close to the measured mean values. To verify whether there is a statistical difference between the estimated and measured mean of distress values, pairwise $t$ tests were conducted. The null hypothesis ($H_0$) was that there is no obvious difference between measured and estimated distress values, and the alternative hypothesis ($H_1$) was that there is a significant difference between them. We do not reject the null hypothesis at the 5% level of significance for all distress indicators. Thus, it can concluded that there is no significant difference between the estimated distress values obtained from the models and the measured distress values for all distress indicators, for both AC and JPCP.

Table 8. Paired Comparisons between Measured and Estimated Distress Indicators

| Model | Mean | | Paired comparison ($\alpha = 0.05$) | |
| --- | --- | --- | --- | --- |
| | Measured | Estimated | $p$-Value | Test result |

| | | | | | |
|---|---|---|---|---|---|
| AC pavements | IRI | 77.78 inch/mile | 78.74 inch/mile | 0.47 | Do not reject $H_0$ |
| | PDI | 3.63 | 3.63 | 0.93 | Do not reject $H_0$ |
| | PQI | 3.43 | 3.42 | 0.85 | Do not reject $H_0$ |
| | PSI | 3.68 | 3.65 | 0.18 | Do not reject $H_0$ |
| JPCP pavements | IRI | 102.33 inch/mile | 102.37 inch/mile | 0.97 | Do not reject $H_0$ |
| | PDI | 4.66 | 4.66 | 0.68 | Do not reject $H_0$ |
| | PQI | 4.09 | 4.09 | 0.98 | Do not reject $H_0$ |
| | PSI | 3.34 | 3.35 | 0.42 | Do not reject $H_0$ |

6.2 Model Limitations

Similar to any other pavement performance evaluation model, the model developed in this paper is only an approximation of the actual physical phenomenon of pavement performance. However, the estimated and measured distress indicators show the applicability of the developed models for pavement performance evaluation in SC. There were still some limitations to the size of the data sample, which might have occurred a few not-so-accurate estimates, especially in the case of PDI in JPCP. A possible approach to overcome this limitation would be to obtain more pavement performance data so that the models could be updated. Furthermore, data from other available sources could be considered. By doing so, in addition to have a relatively large data set, a few more new variables could also be incorporated in the evaluation models, which may help to reduce potential bias and estimation error. However, one should apply engineering judgement before investigating additional variables for the models. Another limitation of the developed models is that they are of an empirical nature. That means the applicability of them outside of SC is limited. To be used for areas outside of SC, it is required to develop calibration process for the models.

## 7. Discussions and Conclusions

The following conclusions can be made based on the analyses of performance data from in-service pavements in SC.

- For the IRI models developed for AC and JPCP, AADT, FFS and precipitation showed statistically significant effects on IRI ($p < 0.01$). AADT and precipitation have positive effects on IRI, whereas FFS has negative effects on IRI for both AC and JPCP. That means IRI increases with increasing AADT and increasing precipitation. However, IRI decreases with increasing FFS.
- For the PSI models developed for AC and JPCP, AADT, FFS and precipitation showed statistically significant effects on PSI ($p < 0.01$). AADT and precipitation have negative effects on PSI, whereas FFS has positive effects on PSI for both AC and JPCP. That means PSI decreases with increasing AADT and increasing precipitation. However, PSI increases with increasing FFS. As PSI is a function of only IRI, different models for these two dependent variables showed similar results.
- Temperature does not show any significant effect on IRI or PSI. Temperature showed significant effect only on AC pavement PDI and PQI ($p < 0.01$). Moreover, positive effect of temperature on AC pavement PDI and PQI means these indices increase with increasing temperature.
- Precipitation was found to be a significant predictor for PSI on both types of pavement, JPCP PDI and PQI, and AC and JPCP IRI ($p < 0.05$). Therefore, the climate input precipitation was found to be more important than temperature for predicting different pavement performance in SC.
- Considering soil type, Type B soil produced statistically higher PDI and PQI ($p < 0.01$) compared to Type A soil on AC pavements; whereas, Type B soil produced statistically higher IRI and PSI ($p < 0.001$) compared to Type A soil on JPCP. Therefore, effects of soil types on pavement performance should be further investigated by performing in-situ tests of subgrade strength.
- Temperature, FFS and soil type showed statistically significant effect on AC pavement PDI and PQI. However, for JPCP PDI and PQI, AADT and precipitation were found as significant variables. AC pavements and JPCP showed different significant variables for PDI and PQI. AADT did not show any significant effects on AC pavement PDI and PQI. Very low regression coefficients obtained from AC pavement PDI model (Adjusted $R^2$ = 0.128) and AC pavement PQI model (Adjusted $R^2$ = 0.152). This result was not expected and may be a result of routine maintenance performed by the SCDOT on AC pavements which was not considered in this study and/or there might be some inconsistencies in the SCDOT's manual survey techniques for distress measurements.
- In this study, performance evaluation models were developed for the SC pavements, taking into account both local and MEPDG distress indices. Therefore, using the developed performance evaluation models, different local pavement performance indicators for a given climatic (temperature and precipitation), traffic (AADT) and material (soil type A or B, pavement type AC or JPCP) condition can be predicted. Additionally, developed performance evaluation models for IRI would also be a useful tool for MEPDG local calibration, to predict IRI in different climatic, traffic and material conditions.

## 8. Future Studies

One of the key findings of this study is that two different types of soil have statistically significant effects on SC pavement performance. Therefore, the difference in their subgrade strength needs to be further investigated. In future studies, resilient modulus of subgrade soil would be determined for both Type A and Type B and the results would be compared. Moreover, precipitation might be considered as a more important climate input than temperature in MEPDG local calibration for SC. Furthermore, the IRI models of this study would be compared with the IRI models developed using MEPDG for both AC and JPCP to better understand the roughness behavior of SC pavements.

**Acknowledgement**
This paper is based on a research supported by the SCDOT and the Federal Highway Administration (FHWA) under contract SPR 708: Calibration of the AASHTO Pavement Design Guide to SC Conditions – Phase I. The authors would like to acknowledge the SCDOT for providing the pavement performance and traffic data. The authors would like to give special thanks to Mr. Jesse Thomson, Jr. and Mr. Chad Rawls for assisting with the study. The opinions, findings and conclusions expressed herein are those of the authors and not necessarily those of the SCDOT or FHWA.

**References**

American Association of State Highway and Transportation Officials (AASHTO) (2008). *Standard Specification for Classification of Soils and Soil-Aggregate Mixtures for Highway Construction Purposes*, M 145-91, Washington, D. C.

Ahmed, K., Abu-Lebdeh, G., and Lyles, R. W. (2006). "Prediction of Pavement Distress Index with Limited Data on Causal Factors: An Auto-Regression Approach." *International Journal of Pavement Engineering*, Vol. 7, No. 1, pp. 23–35, DOI: 10.1080/10298430500502017.

Al-Mansour, A., Sinha, K. C., and Kuczek, T. (1994). "Effects of Routine Maintenance on Flexible Pavement Condition." *Journal of Transportation Engineering*, Vol. 120, No. 1, pp. 65–73, DOI: 10.1061/(ASCE)0733-947X(1994)120:1(65).

ARA, Inc. (2004). *Guide for Mechanistic-Empirical Design of New and Rehabilitated Pavement Structures*, National Cooperative Highway Research Program, Report 1-37A.

Archilla, A. R., and Madanat, S. (2001). "Statistical Model of Pavement Rutting in Asphalt Concrete Mixes." *Transportation Research Record: Journal of the Transportation Research Board*, No. 1764, pp. 70–77, DOI: 10.3141/1764-08.

Baus, R. L., and Hong, W. (2004). *Development of Profiler-Based Rideability Specifications for Asphalt Pavements and Asphalt Overlays*, Federal Highway Administration, Report GT04-07.

Baus, R. L., and Stires, N. R. (2010). *Mechanistic-Empirical Pavement Design Guide Implementation*, Federal Highway Administration, Report GT006-10.

Chan, P. K., Oppermann, M. C., and Wu, S.-S. (1997). "North Carolina's Experience in Development of Pavement Performance Prediction and Modeling." *Transportation Research Record: Journal of the Transportation Research Board*, No. 1592, pp. 80–88, DOI: 10.3141/1592-10.

Chang, C.-M., Baladi, G. Y., and Wolff, T. F. (2001). "Using Pavement Distress Data to Assess Impact of Construction on Pavement Performance." *Transportation Research Record: Journal of the Transportation Research Board*, No. 1761, pp. 15–25, DOI: 10.3141/1761-03.

Chu, C.-Y., and Durango-Cohen, P. L. (2008). "Empirical Comparison of Statistical Pavement Performance Models." *Journal of Infrastructure Systems*, Vol. 14, No. 2, pp. 138–149, DOI: 10.1061/(ASCE)1076-0342(2008)14:2(138).

Cooper, S.B., Elseifi, M.A., and Mohammad, L.N. (2012). "Parametric Evaluation of Design Input Parameters on the Mechanistic-Empirical Pavement Design Guide Predicted Performance." *International Journal of Pavement Research and Technology*, Vol. 5, No. 4, pp. 218–224.

DeLisle, R. R., Sullo, P., and Grivas, D. A. (2003). "Network-Level Pavement Performance Prediction Model Incorporating Censored Data." *Transportation Research Record: Journal of the Transportation Research Board*, No. 1853, pp. 72–79, DOI: 10.3141/1853-09.

Dowling, R., Kittelson, W., Zegeer, J., and Skabardonis, A. (1997). *Planning Techniques to Estimate Speeds and Service Volumes for Planning Applications*, National Cooperative Highway Research Program, Report 387.

Ferreira, A., Picado-Santos, L. D., Wu, Z., and Flintsch, G. (2011). "Selection of Pavement Performance Models for use in the Portuguese PMS." *International Journal of Pavement Engineering*, Vol. 12, No. 1, pp. 87–97, DOI: 10.1080/10298436.2010.506538.

Federal Highway Administration (FHWA) (2004). *Pavement Smoothness Methodologies*, FHWA-HRT-04-061 145-91. <www.fhwa.dot.gov/pavement/smoothness/index.cfm>

Gulen, S., Zhu, K., Weaver, J., Shan, J., and Flora, W. F. (2001). *Development of Improved Pavement Performance Prediction Models for the Indiana Pavement Management System*, Federal Highway Administration, Report FHWA/IN/JTRP-2001/17.

Gupta, A., Kumar, P., and Rastogi, R. (2014). "Critical Review of Flexible Pavement Performance Models." *KSCE Journal of Civil Engineering*, Vol. 18, No. 1, pp. 142–148, DOI: 10.1007/s12205-014-0255-2.

Hasan, M. R. M., Hiller, J. E., and You, Z. (2015). "Effects of Mean Annual Temperature and Mean Annual Precipitation on the Performance of Flexible Pavement using ME Design." *International Journal of Pavement Engineering* (forthcoming), DOI: 10.1080/10298436.2015.1019504.
18

Henning, T. F. P., Costello, S. B., Dunn, R. C. M., Parkman, C. C., and Hart, G. (2004). "The Establishment of A Long-Term Pavement Performance Study on the New Zealand State Highway Network." *Road and Transport Research*, Vol. 13, No. 2, pp. 17–32.

Hong, H. P., and Wang, S. S. (2003). "Stochastic Modeling of Pavement Performance." *International Journal of Pavement Engineering*, Vol. 4, No. 4, pp. 235–243, DOI: 10.1080/10298430410001672246.

Isa, A. H. M., Ma'soem, D. M., and Hwa, L. T. (2005). "Pavement Performance Model for Federal Roads." *Proceedings of the Eastern Asia Society for Transportation Studies*, Vol. 5, pp. 428–440.

Johnson, K. D., and Cation, K. A. (1992). "Performance Prediction Development using Three Indexes for North Dakota Pavement Management System." *Transportation Research Record: Journal of the Transportation Research Board*, No. 1344, pp. 22–30.

Keith, T. Z. (2015). *Multiple Regression and Beyond: An Introduction to Multiple Regression and Structural Equation Modeling*, Routledge, Taylor and Francis, New York, USA.

Kim, S.-H., and Kim, N. (2006). "Development of Performance Prediction Models in Flexible Pavement using Regression Analysis Method." *KSCE Journal of Civil Engineering*, Vol. 10, No. 2, pp. 91–96, DOI: 10.1007/BF02823926.

Li, X. Y., Zhang, R., Zhao, X., and Wang, H. N. (2014). "Sensitivity Analysis of Flexible Pavements Parameters by Mechanistic-Empirical Design Guide." *Applied Mechanics and Materials*, Vol. 590, pp. 539–545, DOI: 10.4028/www.scientific.net/AMM.590.539.

Lu, D. Y., Lytton, R. L., and Moore, W. M. (1974). *Forecasting Serviceability Loss of Flexible Pavements*, Federal Highway Administration, Report TTI-2-8-74-57-1F.

Madanat, S. (1993). "Incorporating Inspection Decisions in Pavement Management." *Transportation Research Part B: Methodological*, Vol. 27, No. 6, pp. 425–438, DOI: 10.1016/0191-2615(93)90015-3.

Meegoda, J. N., and Gao, S. (2014). "Roughness Progression Model for Asphalt Pavements using Long-Term Pavement Performance Data." *Journal of Transportation Engineering*, Vol. 140, No. 8, pp. 1–7, DOI: 10.1061/(ASCE)TE.1943-5436.0000682.

Mikhail, M. Y., Mamlouk, M. S. (1999). "Effect of Traffic Load on Pavement Serviceability." *ASTM Special Technical Publication*, No. 1348, pp 7–20.

Mills, L. N. O., Attoh-Okine, N. O., and McNeil, S. (2012). "Developing Pavement Performance Models for Delaware." *Transportation Research Record: Journal of the Transportation Research Board*, No. 2304, pp. 97–103, DOI: 10.3141/2304-11.

Nassiri, S., and Bayat, A. (2013). "Evaluation of MEPDG Seasonal Adjustment Factors for the Unbound Layers' Moduli using Field Moisture and Temperature Data." *International Journal of Pavement Research and Technology*, Vol. 6, No. 1, pp. 45–51.

National Centers for Environmental Information (NCEI) (2015). *National Oceanic and Atmospheric Administration.* <www.ncdc.noaa.gov>

Neter, J., and Wasserman, W. (1996). *Applied Linear Statistical Models*. Irwin, Chicago.

Orobio, A., and Zaniewski, J.P. (2011). "Sampling-Based Sensitivity Analysis of the Mechanistic-Empirical Pavement Design Guide Applied to Material Inputs." *Transportation Research Record: Journal of the Transportation Research Board*, No. 2226, pp. 85–93, DOI: 10.3141/2226-09.

Paterson, W. D. O. (1986). "International Roughness Index: Relationship to other Measures of Roughness and Riding Quality." *Transportation Research Record: Journal of the Transportation Research Board*, No. 1084, pp. 49–59.

Pierce, C. E., Gassman, S. L., and Ray, R. P. (2011). *Geotechnical Materials Database for Embankment Design and Construction*, Federal Highway Administration, Report FHWA-SC-11-02.

PMS Inc. (1990). "PMS Final Specification Report." South Carolina Department of Transportation, Columbia, SC.
19